Costantino Sigismondi
*Università di Roma
"La Sapienza"*
sigismondi@icra.it

# Misure quantitative del seeing atmosferico

**Abstract:** A simple technique of measurement of seeing in daytime and nighttime, based upon drift-scan observations, is presented, along with observational examples. This experience can be repeated easily in classroom contexts.

**Introduzione:** Come vedere la macchia rossa…o la *nix olympica* con le osservazioni al telescopio ad occhio nudo? Solitamente a questa domanda si risponde in termini di potere risolutivo del telescopio, e dunque di diametro dell'obbiettivo. Tuttavia maturando un po' di esperienza osservativa si capisce che quanto sia importante aspettare un momento di calma nel seeing per apprezzare i dettagli più minuti, sfruttare al meglio il potere risolutivo del telescopio.

Nel diario delle osservazioni si usa annotare il seeing seguendo la scala di Antoniadi, raccomandata dalla sezione pianeti dell'UAI e dell'ALPO.

Oggi le tecniche di ripresa con CCD hanno permesso di raggiungere risultati che solo dieci anni fa erano appannaggio esclusivo degli osservatori professionali.

Oltre alla scala di Antoniadi [1], sempre valida soprattutto didatticamente, è possibile associare un valore del seeing misurato con un video drift-scan, che segua le riprese dedicate all'imaging. Si propone di seguito un metodo, con particolare applicazione alle osservazioni del Sole.

**Ruolo del seeing nelle osservazioni solari**

Nel nuovo ciclo di attività solare, il XXIV che sembra il preludio di una fase di grande minimo tipo Maunder, le rarissime macchie sono piccole e di durata piuttosto effimera, si formano e si dissolvono in poche ore. Una valutazione visuale sul seeing è più difficile mancando le macchie, perciò è opportuno fare una misura col metodo del drift-scan sfruttando il lembo solare.

Nelle misure di diametro solare, infine, il seeing gioca un ruolo chiave. In presenza di turbolenza il bordo del Sole, definito come il punto dove è il massimo della derivata del profilo di luminosità perpendicolare al lembo, si sposta verso l'esterno, dando luogo ad un aumento spurio del diametro misurato. Pur esistendo modelli teorici [2] per questo fenomeno, non esiste ancora un atlante di dati osservativi che mettano in correlazione i diversi valori del seeing istantaneo ed il diametro misurato. La realizzazione di questo atlante è prevista nel progetto CLAVIUS da condursi al telescopio solare di Locarno (Svizzera Italiana, Canton Ticino). Misure amatoriali potrebbero contribuire con i primi punti sperimentali di questo atlante.

**Le varie componenti del seeing**

Entrare nel dettaglio dell'ottica atmosferica esula dallo scopo di questo articolo, tuttavia è bene avere un'idea dei fenomeni e delle loro cause anche parziale.

L'Image motion consiste nello spostamento del baricentro dell'immagine, solitamente stellare o planetaria, ma il discorso è valido anche per il Sole.

Il Blurring invece è la deformazione dell'immagine mentre il baricentro resta fisso.

Le frequenze più alte in gioco sono attorno ai 100 Hz, mentre l'image motion può agire anche su scale temporali superiori al secondo.

Il parametro di Fried $r_0$ corrisponde al diametro del telescopio il cui limite di diffrazione corrisponde al valore del seeing in quel momento.

Dunque $2.44 \cdot 10^5 \cdot \lambda / r_0 = \varrho"$,

dove $\varrho"$ è il seeing in secondi d'arco e $\lambda$ e $r_0$ sono misurati in metri.

Avere un telescopio con diametro molto maggiore del parametro di Fried determina la formazione di speckles, macchie della stessa sorgente sparse qua e là su piano focale, richiedendo complicate procedure di analisi delle immagini per risalire all'immagine imperturbata. Sono ottimi valori del parametro di Fried $r_0=20$ cm corrispondenti a $\varrho=0.5"$. Ma in genere di giorno i valori tipici per il parametro di Fried sono $r_0<7$cm, e di notte $r_0<10$cm.

Nei modelli realistici di seeing entrano in gioco almeno quattro parametri: l'altezza scala superiore ed inferiore a cui si verificano i fenomeni di turbolenza, la velocità dell'aria e la dimensione delle celle di turbolenza.

Se abbiamo turbolenza nelle vicinanze del telescopio, come di giorno quando il suolo è scaldato direttamente dal Sole che solleva colonne convettive d'aria, la componente di image motion è maggiore poiché le celle di turbolenza sono vicine al telescopio. Un altro caso evidente di turbolenza nelle vicinanze del telescopio è la presenza di alberi rispetto ai quali il telescopio è sottovento. Basta spostarsi sopravento rispetto agli alberi o agli spigoli dei palazzi per limitare di molto la turbolenza.

Anche la copertura dell'osservatorio può giocare un ruolo chiave nella nitidezza delle immagini: le cupole tradizionali offrono un profilo che rompe i flussi d'aria che sono nelle vicinanze del telescopio, creando turbolenze locali. I moderni osservatori stanno abbandonando la tradizionale copertura a cupola, sostituendola con tetti a scorrimento, dai quali il telescopio emerge o è comunque protetto da flussi turbolenti.

**Misura del seeing diurno con il metodo del drift-scan**

Si punta il Sole e poi si blocca l'inseguimento motorizzato del telescopio, il moto in ascensione retta ovvero l'inseguimento altazimutale delle montature più moderne. Il drift è la deriva dell'immagine, per l'effetto della rotazione terrestre [3], che può essere proiettata su uno schermo con una griglia regolare stampata sopra [4]. Si può far stampare su un foglio bianco una tabella di Word, che va benissimo allo scopo. Questa immagine in movimento viene ripresa con una videocamera. Gli intertempi misurati al passaggio del bordo dell'immagine del Sole sulle varie righe dovrebbero essere tutti uguali in assenza di agitazione atmosferica, invece la loro deviazione standard σ non è nulla.

Il seeing è, in prima approssimazione, dato dalla formula
ρ["]≈15"·σ[s]·cosδ,
dove σ è la deviazione standard misurata in secondi delle differenze –intertempi- di contatto su una griglia omogenea, e δ è la declinazione del Sole al momento dell'osservazione.
Dai dati di un drift-scan al telescopio Gregory-Coudé di Locarno (Svizzera) ripreso il 9 agosto 2008 con un videocamera SANYO CG9 a 60 fps, si è ottenuta la seguente tabella.

| Intertempi [s] | Transiti su righe parallele [s] |
|---|---|
|  | 4.3 |
| 0.3 | 4.6 |
| 0.27 | 4.87 |
| 0.37 | 5.24 |
| 0.25 | 5.49 |
| 0.32 | 5.81 |
| 0.31 | 6.12 |
| 0.27 | 6.39 |
| 0.32 | 6.71 |
| 0.4 | 7.11 |
| 0.3 | 7.41 |
| 0.3 | 7.71 |
| 0.28 | 7.99 |
| 0.28 | 8.27 |
| Deviazione standard [s] | Seeing ["] |
| σ=0.041 s | ρ=0.60" |

Il giorno 9 agosto 2008 alle 15 UT il Sole si trovava alla declinazione δ=15°38', dunque alla deviazione standard dei dati in prima colonna, σ=0.041 s, corrisponde un seeing di ρ=0.60", decisamente un ottimo seeing. La granulazione era ben visibile, tuttavia nessuna macchia appariva sul Sole quel giorno.

A Locarno si è già verificato che il seeing migliora nel pomeriggio, laddove usualmente questo è migliore all'alba, quando terra ed aria raggiungono la stessa temperatura, riducendo al minimo la turbolenza. La presenza delle acque del Lago Maggiore nei settori a Sud e a Sud-Ovest dell'Osservatorio determina circostanze ambientali particolari, che favoriscono ottimi seeing di pomeriggio.

Questo metodo può essere applicato tale e quale nelle osservazioni stellari o planetarie, lasciando transitare sui pixel del rivelatore il corpo celeste al massimo ingrandimento.

**Seeing notturno**

Un video di Fomalhaut è stato ripreso l'8 ottobre 2009 alle 0:55 locali da Roma, con videocamera SANYO HD1010 a 60 fps. Questa risoluzione temporale consentirebbe un'accuratezza di 1/60 s ovvero meno di 0.02 s, nella determinazione degli istanti in cui l'immagine della stella tocca la griglia di riferimento. Il video in formato Mpeg4 è fatto scorrere fotogramma per fotogramma con il programma Quicktime 7 con la finestra che mostra le impostazioni del fimato aperta per leggere il tempo corrispondente a ciascun fotogramma. L'immagine (con coma) della stella si deforma e cambia di intensità ad ogni fotogramma. Nelle sequenza di immagini seguente si vedono 5 fotogrammi consecutivi di Fomalhaut, intervallati di 16 millisecondi l'uno dall'altro.

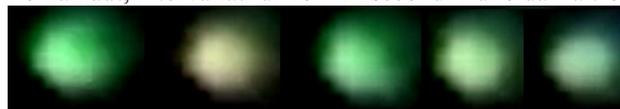

Fig. 1 Sequenza di immagini consecutive di Fomalhaut a 13° sull'orizzonte, ad 1/60 s l'una dall'altra.

La ripresa è stata fatta con zoom digitale 100x per massimizzare la dimensione dell'immagine; la videocamera ne ha uno ottico da 10x.

La seguente tabella riassume delle misure prese dal video di Fomalhaut (δ=-29°35') quando si trovava a 13°17'sull'orizzonte.

| Deviazione standard delle differenze [s] |
|---|
| 0.156 -10 dati |
| 0.135 -10 dati |
| 0.183-10 dati |
| 0.120- 5 dati |
| Media e sigma |
| 0.149±0.027 [s] |
| ρ=1.94"±0.36" |

La griglia di riferimento è stata realizzata con un foglio a quadretti fissato sullo schermo del computer. L'immagine della stella si vede attraverso il foglio, e quando il bordo della stella risulta tangente ad una riga verticale si trascrive il tempo letto sulla finestra di controllo. Si trascrive solo il primo istante in cui questa tangenza si osserva, perché questo può ripetersi varie volte nel corso dei fotogrammi successivi. Lo stesso discorso si ripete con la linea verticale successiva e così di seguito.

La massima risoluzione di questo metodo è quella del singolo fotogramma[0.0166 s]. Campionando nello stesso video della durata di 21 secondi altre serie di tempi di contatto si ottiene la tabella precedente. La media delle deviazioni standard corrisponde a σ=0.15±0.03 s, è stata tradotta in secondi d'arco mediante l'equazione
ρ["]=σ[s]·15.04"/s·cos(δ)
valida per le stelle fisse.

**Referenze**